\documentclass[superscriptaddress,amsmath,amssymb,aps]{revtex4-2}

\usepackage{graphicx}
\usepackage{dcolumn}
\usepackage{bm}

\begin{document}

\title{The Fitness Value of Information with Delayed Phenotype Switching: Optimal Performance with Imperfect Sensing}

\author{Alexander S. Moffett}
\affiliation{Department of Electrical Engineering and Computer Science, York University, Toronto, Ontario, Canada}
\author{Nigel Wallbridge}
\affiliation{Vivent SaRL, Crans-pr{\`e}s-C{\'e}ligny, Switzerland}
\author{Carrol Plummer}
\affiliation{Vivent SaRL, Crans-pr{\`e}s-C{\'e}ligny, Switzerland}
\author{Andrew W. Eckford}
\email{aeckford@yorku.ca}
\affiliation{Department of Electrical Engineering and Computer Science, York University, Toronto, Ontario, Canada}

\begin{abstract}
The ability of organisms to accurately sense their environment and respond accordingly is critical for evolutionary success. However, exactly how the sensory ability influences fitness is a topic of active research, while the necessity of a time delay between when unreliable environmental cues are sensed and when organisms can mount a response has yet to be explored at any length. Accounting for this delay in phenotype response in models of population growth,
we find that a critical error probability can exist under certain environmental conditions: an organism with a sensory system with any error probability less than the critical value can achieve the same long-term growth rate as an organism with a perfect sensing system. We also observe a trade off between the evolutionary value of sensory information and robustness to error, mediated by the rate at which the phenotype distribution relaxes to steady-state. The existence of the critical error probability could have several important evolutionary consequences, primarily that sensory systems operating at the non-zero critical error probability may be evolutionarily optimal. 
\end{abstract}

\maketitle


\section{Introduction}

It has long been clear that the concept of information should be of great importance in biology \cite{rashevsky1950some,quastler1953information,johnson1970information,smith2000concept,tkacik2016information}. Information is invoked perhaps most commonly 
in reference to the central dogma of molecular biology. Research in neuroscience \cite{dimitrov2011information}, cell biology \cite{cheong2011information,rhee2012application}, ecology \cite{harte2014maximum}, and evolutionary biology \cite{donaldson2010fitness,rivoire2011value} have also benefited from both formal and informal use of information in explaining various phenomena therein. However, the appropriate definitions and interpretations of information are not as clear in biological contexts as they are in the context of human communication, the application for which information theory was originally constructed \cite{shannon1948mathematical}. One intriguing line of inquiry seeks to understand the role of information in evolution, with the central question: what is the fitness value of information? In other words, how does the information gathered by the sensory systems of a type of organism relate to its fitness?  

Recent theoretical work has begun to explore the connections between information and fitness, building on the work of Kelly \cite{kelly1956new} concerning optimal betting strategies, for example on horse races, when unreliable information concerning the outcomes of races is available. Kelly considered a gambler who can partition their wealth into bets on each horse in a given race. The gambler's goal is to maximize the long-term growth rate of their wealth based on unreliable information, called side information, about the race's outcome. By unreliable information, we mean that the probability that the gambler will be fed an erroneous outcome is greater than zero, but less than the error using only their prior information about the race. Using the geometric mean in his growth rate definition, Kelly found that, in certain conditions, the log of the optimal long term growth rate with side information minus the log of the optimal growth rate with no side information, is equal to the mutual information across the side information channel, showing that there is a well-defined value of unreliable information.

Applications of Kelly's work to evolutionary biology utilize the same measure of long-term growth, called the dominant Lyapunov exponent, or simply the Lyapunov exponent. The Lyapunov exponent can be defined \cite{metz1992should} as
\begin{equation}
\Lambda=\lim_{T\rightarrow\infty}\log\bigg(\frac{N_{T}}{N_{0}}\bigg)
\end{equation}
where $N_{T}$ is the total population size at time $T$ and $N_{0}$ is the initial total population size. Instead of a gambler placing bets on horse races, we can consider a population of organisms with the ability to stochastically switch phenotypes. Instead of maximizing the long-term growth of wealth, we can examine the long-term growth of the population, which is partitioned into subpopulations each expressing a certain phenotype. For example, consider a population with $K$ subpopulations ($\phi_{i}$ for $i\in\{1,2,\dots,K\}$) constituting fixed fractions $f(\phi_{i})$ of the overall population such that $\sum_{i=1}^{K}f(\phi_{i})=1$. If the environment can exist in $K$ different states ($e_{j}$ for $j\in\{1,2,\dots,K\}$) such that each subpopulation can have different growth rates $w(\phi_{i},e_{j})$ in different environments $e_{j}$ and successive environmental states are independent, it can be shown that the long-term growth rate is
\begin{equation}
\Lambda=\sum_{j=1}^{K}p(e_{j})\log\bigg(\sum_{i=1}^{K}f(\phi_{i})w(\phi_{i},e_{j})\bigg)
\end{equation}
The Lyapunov exponent is a convenient measure of overall population fitness in situations such as these, with structured populations and time-dependent environmental conditions \cite{metz1992should}. 

Donaldson-Matasci, Bergstrom, and Lachmann \cite{donaldson2010fitness} considered populations of organisms capable of developing into one of several phenotypes, each with a fitness dependent on the state of the environment. Considering a random environment and a noisy channel for environmental sensing, they were able to define a ``fitness value of information'' in terms of the long-term growth rate for optimal probabilistic developmental strategies. Rivoire and Leibler \cite{rivoire2011value} extended this work for a number of more complicated models, including the case of phenotype switching in non-independent and identically distributed (IID) environments and with non-IID phenotypes as described by Kussell and Leibler \cite{kussell2005phenotypic}. A number of more recent studies have further built upon this line of inquiry \cite{xue2018benefits,xue2019environment,tal2020adaptive}. 

In the case of a non-IID environment, an organism can improve its prediction of future environmental states by conditioning on knowledge of current and past states. Alternatively, if successive environmental states are entirely independent, conditioning on present and past environmental states is of no predictive value. In this case, a bet-hedging approach would be optimal, where the strategy is agnostic to the environment at any one time but is tuned according to the probabilities of environmental state occurrence through selection. The situation is further complicated if conditional phenotype expression probabilities are themselves dependent on past phenotypic states \cite{kussell2005phenotypic,mayer2017transitions,hufton2018phenotypic}, so that there is a time period where the distribution of phenotypes must relax to a new distribution upon environmental change \cite{padilla1996plastic}.

\begin{figure}[ht!]
    \centering
    \includegraphics[width=.75\textwidth]{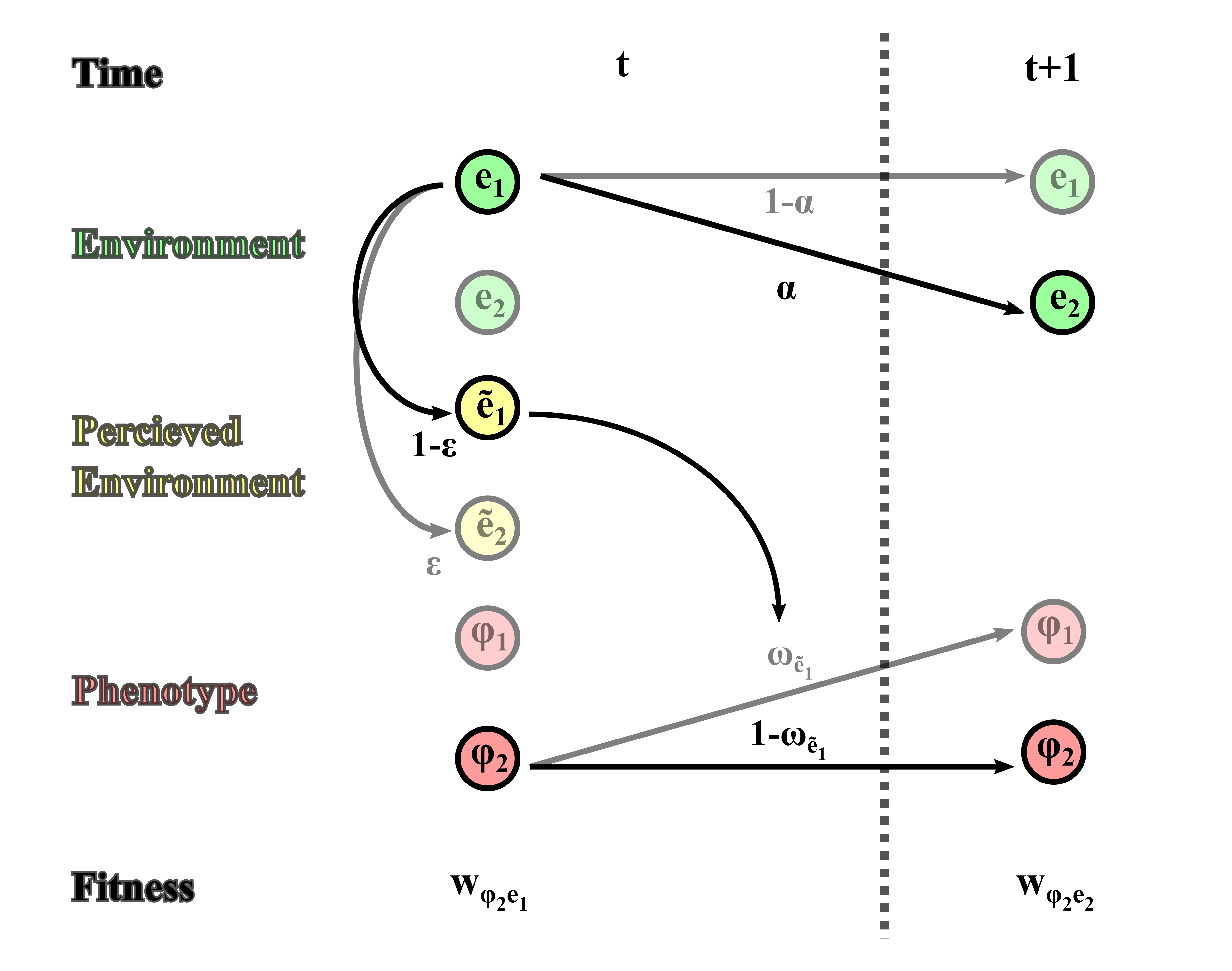}
    \caption{{\bf Model schematic.} This schematic illustrates an example of the process an individual in our model can follow. In this case, the environment is in state $e_{1}$ and the phenotype of the individual is in state $\phi_{2}$ at time $t$. These two variables alone determine the fitness of our individual at time $t$, which is $w_{\phi_{2}e_{1}}$. The environment perceived by our example individual is chosen according to the error probability $\epsilon$. Our example individual correctly perceives the environment, so the perceived environment is $\tilde{e}_{1}$, chosen with probability $1-\epsilon$. The individual then expresses phenotype switching probabilities $\chi_{\tilde{e}_{1}}$ and $\omega_{\tilde{e}_{1}}$ accordingly. However, because the phenotype at time $t$ is $\phi_{2}$ only $\omega_{\tilde{e}_{1}}$ is relevant at the moment. The environmental state for $t+1$ is then chosen according to the environmental switching probabilities $\alpha$ and $\beta$, where $\alpha$ is the relevant value because the environment is in $e_{1}$ at time $t$. Here, state $e_{2}$ is chosen with probability $\alpha$. The organism chooses phenotype $\phi_{2}$ for time $t+1$ according to the probability $1-\omega_{\tilde{e}_{1}}$. The fitness of our individual is now $w_{\phi_{2}e_{2}}$. Then the perceived state can again be chosen and the process repeated. This figure was created using Inkscape 0.92 \protect\cite{inkscape}.}
    \label{fig:schematic}
\end{figure}

We examine the problem of optimal Markov phenotype switching in the presence of a non-IID environment using a two state Markov chain description of environmental dynamics and a two state, environment-dependent Markov chain description of phenotype switching (Fig. \ref{fig:schematic}). This extends the work of Rivoire and Leibler (Appendices D, E, and H therein) \cite{rivoire2011value} on solvable models of optimal population growth in non-IID environments to account for Markov phenotype switching. Our model introduces a delay between the time of environmental sensing and organismal response, meaning that populations must ``plan ahead'' since any response will occur in the future, even with an infinite phenotype switching rate. We examine optimal phenotype switching strategies and the corresponding long-term growth rates as the probability of making an error in sensing the environment goes from zero (perfect information) to 1/2 (no information). We also derive an approximate Lyapunov exponent for the case of a finite phenotype switching rate and numerically explore optimal strategies at different error probabilities and phenotype switching rates. 

Consistent with previous work, we show that the fitness value of information is bounded from above by the mutual information between the environmental state at one time and the immediately following environmental state. We find that below a critical environmental sensing error probability, it is possible for a population to adopt a strategy leading to the same long-term growth rate as when there is zero probability of sensing error, suggesting a trade-off between optimal growth rate and sensory robustness in different environments. In summary, we report the following:

\begin{itemize}
    \item Derivations for approximate Lyapunov exponents describing the long-term growth rates of populations with a delay in response and imperfect environmental sensing.
    \item The fitness value of perfect sensing when the phenotype switching rate approaches infinity is the mutual information between consecutive environmental states.
    \item Subject to given conditions, imperfect sensing has no impact on the fitness value of information below a critical error probability.
    \item Slower phenotype switching rates reduce the fitness value of information while increasing robustness to imperfect sensing.
\end{itemize}

\section{Models}

\subsection{The environment}
We define the environment as all factors excluding the internal state of individuals in the population of interest, excluding those factors that depend on the density of the population. (Including density-dependent environmental factors in our model is a direction for future work.) In order to simplify analysis, we study an environment which can exist in one of two possible states at any given time. We denote the set of environmental states as $\mathcal{E}=\{e_{1},e_{2}\}$. Throughout this paper, we use the notation 
$e'\in\mathcal{E}\setminus\{e\}$, so that $e'$ is the environmental state other than $e$.

We assume that the environment evolves in time as a discrete-time Markov process $\{E_{m}\}_{m=0}^{M}$, meaning that the probability that an environmental state occurs in the next time step depends only on the current environmental state. The transition probability matrix describing this Markov process is
\begin{equation}
\mathbf{R}=
\begin{bmatrix}
    1-\alpha & \beta \\
    \alpha & 1-\beta
\end{bmatrix}
\end{equation}
where $\alpha$ is the probability that the environment will switch from state $e_{1}$ to $e_{2}$ in a single time step and $\beta$ is the probability of the inverse process. We assume that the environment is persistent, so that $\alpha,\beta\in(0,1/2)$. The stationary probabilities of each environmental state are $p(e_{1})=\frac{\beta}{\alpha+\beta}$ and $p(e_{2})=\frac{\alpha}{\alpha+\beta}$. We write the mean lifetime of each environmental state as $\mu_{e_{1}}=1/\alpha$ and $\mu_{e_{2}}=1/\beta$ and the overall mean environmental state lifetime as $\mu=\frac{1}{2}(\mu_{e_{1}}+\mu_{e_{2}})$.

\subsection{Phenotypes and phenotype switching}

We consider a large, genetically homogeneous population of organisms growing in the environment. To keep analysis simple, individuals can exist in one of two phenotypic states. We write the set of phenotypic states as $\Phi=\{\phi_{1},\phi_{2}\}$. 

We assume that the phenotype of an individual can change stochastically according to a time-inhomogeneous (meaning time-dependent) discrete-time Markov chain. Because we are interested in the role of environment sensing in fitness, the phenotype transition probabilities depend on the \emph{perceived} state of the environment $\tilde{e}$. By the perceived environmental state, we mean the environmental state that an individual perceives through some sensory mechanism. If the sensory mechanism is error-prone, the perceived environmental state may not be the same as the actual environmental state. Because the environment, and therefore the perceived environment, changes with time, the phenotype transition probabilities also depend on time. The probability that the phenotype of an individual will switch from one state to another can be arranged in matrix form
\begin{equation}
\mathbf{P}_{\tilde{e}}=
\begin{bmatrix} 
	1-\chi_{\tilde{e}} & \omega_{\tilde{e}} \\
	\chi_{\tilde{e}} & 1-\omega_{\tilde{e}} \\
\end{bmatrix}
\end{equation}
given that the perceived environment is in state $\tilde{e}\in\mathcal{E}$. We do not require phenotypes to be persistent, so the only restrictions on $\chi_{\tilde{e}_{1}}$, $\omega_{\tilde{e}_{1}}$, $\chi_{\tilde{e}_{2}}$, and $\omega_{\tilde{e}_{2}}$ are that they fall between zero and one, inclusive, and that $\chi_{\tilde{e}_{1}}+\omega_{\tilde{e}_{1}},\chi_{\tilde{e}_{2}}+\omega_{\tilde{e}_{2}}<1$.

It is important to emphasize that because the population is genetically homogenous, these phenotypes represent different states that an individual can exist in over its lifetime, and reversibly switch between. We assume that the phenotypic state of an individual is not heritable, so that an individual of one phenotype can have offspring displaying either phenotype.

We denote the normalized right (column) eigenvector of $\mathbf{P}_{\tilde{e}}$ corresponding to an eigenvalue of $\lambda^{(1)}_{\tilde{e}}=1$ as $\bm{\pi}_{\tilde{e}}$
\begin{equation}\label{eq:phenotype_column}
\bm{\pi}_{\tilde{e}}=
\begin{bmatrix}
    \pi_{\phi_{1},\tilde{e}} \\
    \pi_{\phi_{2},\tilde{e}}
\end{bmatrix}
\end{equation}
which describes the steady-state distribution of phenotypes in perceived environmental state $\tilde{e}$. We denote the left (row) and right (column) eigenvectors of $\mathbf{P}_{\tilde{e}}$ corresponding to the eigenvalue $\lambda^{(2)}_{\tilde{e}}=1-\chi_{\tilde{e}}-\omega_{\tilde{e}}$ as $\mathbf{l}_{\tilde{e}}$ and $\mathbf{v}_{\tilde{e}}$, respectively. In a constant perceived environment, the rate at which the phenotype distribution decays to a steady state is given by $k_{\tilde{e}}=-\log(\lambda^{(2)}_{\tilde{e}})$, so that this geometric decay can be represented as $\exp(-k_{\tilde{e}}n)$ at time step $n$. We assume that this rate is the same in both perceived environments, that is $k_{\tilde{e}_{1}}=k_{\tilde{e}_{2}}=k$ (see Appendix \ref{section:appendixb} for the case where $k_{\tilde{e}_{1}}\neq{}k_{\tilde{e}_{2}}=k$). This implies that $\chi_{\tilde{e}_{1}}+\omega_{\tilde{e}_{1}}=\chi_{\tilde{e}_{2}}+\omega_{\tilde{e}_{2}}$, and that the decay rates for the overall environments are also equal ($k_{e_{1}}=k_{e_{2}}$, see Appendix \ref{section:appendixb} for derivations of $k_{e_{1}}$ and $k_{e_{2}}$).

\subsection{Fitness}

We write the fitness for each phenotype in environmental state $e$ as a vector
\begin{equation}\label{eq:fitness_column}
\mathbf{w}_{e}=
\begin{bmatrix}
    w_{\phi_{1},e} \\
    w_{\phi_{2},e}
\end{bmatrix}
\end{equation}
where we require, without loss of generality, that $w_{\phi_{1},e_{1}}>w_{\phi_{2},e_{1}}$ and $w_{\phi_{1},e_{2}}<w_{\phi_{2},e_{2}}$. These inequalities imply that phenotype $\phi_{1}$ is better suited for environment $e_{1}$ in terms of fitness than phenotype $\phi_{2}$, while phenotype $\phi_{2}$ is better suited for environment $e_{2}$. We occasionally refer to the better-suited phenotype as the ``correct'' phenotype for a given environment and the worse-suited phenotype as the ``incorrect'' phenotype. This is not intended to imply that any probability of expression of an incorrect phenotype leads to a sub-optimal growth rate, which is often untrue as will become clear later on. The case where $w_{\phi_{1},e}=w_{\phi_{2},e}$ is not of interest given the focus of this work, as there would effectively be a single phenotype present in the population where selection is concerned. We define the difference between the better-adapted and worse-adapted phenotypes for an environment as $\Delta{}w_{e_{1}}\equiv{}w_{\phi_{1},e_{1}}-w_{\phi_{2},e_{1}}$ and $\Delta{}w_{e_{2}}\equiv{}w_{\phi_{2},e_{2}}-w_{\phi_{1},e_{2}}$. The fitness values in Eq. \ref{eq:fitness_column} represent a row of a fitness matrix, arranged as
\begin{equation}\label{eq:fitness_matrix}
\mathbf{W}=
\begin{bmatrix}
    w_{\phi_{1}e_{1}} & w_{\phi_{2}e_{1}} \\
    w_{\phi_{1}e_{2}} & w_{\phi_{2}e_{2}}
\end{bmatrix}.
\end{equation}
Because we have assumed that the phenotype of a parent is not passed down to its offspring, we can describe the mean population fitness in terms of $\mathbf{w}_{e}$ and the eigenvalues and eigenvectors of $\mathbf{P}_{\tilde{e}}$. For example, if the environment is independent of time, remaining in a single state $e\in\mathcal{E}$ at all times, and each individual can perfectly sense the environment, the mean population fitness is $\overline{w}(e)=\sum_{\phi\in\Phi}w_{\phi,e}\pi_{\phi,e}=\mathbf{w}_{e}^{T}\bm{\pi}_{e}$. While the stationary phenotype distribution depends on the perceived environmental state rather than the actual environmental state, we use the notation $\bm{\pi}_{e}$ here and throughout this paper to allow for consistency of environmental state indices. 

\subsection{Environmental sensing}

We represent the ability of individuals to sense the environmental state as a distribution $q(\tilde{e}|e)$, the probability of perceiving environmental state $\tilde{e}\in\mathcal{E}$ given that the actual environmental state is $e$. We assume that sensing ability is independent of phenotype, and write the error probability as $q(\tilde{e}|e)=\epsilon$ for $\tilde{e}\neq{}e$ and the probability of correctly perceiving the environmental state as $q(\tilde{e}|e)=1-\epsilon$ for $\tilde{e}=e$.

When sensing is imperfect (when $0<\epsilon<1$) the perceived environmental state will fluctuate rapidly even with a slowly changing environment. When the environment is constant, the effective environment is independent and identically distributed. However, this is not an issue for our analysis as the population will change between two population structures based on effective strategies (see Results, subsection C, and Appendix \ref{section:appendixb}).

\subsection{Population growth}

We use a simple, discrete-time model of population growth, ignoring any limitations on growth imposed by population density. If the environmental state is constant (for example, in state $e_{1}$) and only one phenotype is present (for example, in state $\phi_{1}$), growth occurs according to 
\begin{equation}
N_{M}=(w_{\phi_{1}e_{1}})^{M-1}N_{0}.
\end{equation}
$N_{M}$ is the population size at time step $M$, while $N_{0}$ is the initial population size. In this case, population growth is geometric. If the environment varies with time and multiple phenotypes can exist in the population, then the population grows according to
\begin{equation}\label{eq:growth}
N_{M}=\prod_{m=0}^{M-1}\overline{w}(m,e(m))N_{0}
\end{equation}
where $\overline{w}(m,e(m))$ indicates that the mean population fitness depends explicitly on time and on the time-dependent environmental state.  

In order to account for causality, we assume that there is a single time step delay between when the next environmental state is chosen and when an individual can sense a change (or lack thereof) and respond accordingly. This accounts for the time needed for biochemical or neural processes underlying information processing and decision making to take place.

\section{Results}

\subsection{Approximating the Lyapunov exponent}\label{section:resultsa}

In order to define the fitness value of information, we need expressions for the Lyapunov exponents. These expressions need to be in terms of environment-dependent phenotype distributions, environmental parameters, and the error probability of environmental sensing. In keeping with conventions, we write Lyapunov exponents as $\Lambda$. 

The fitness value of information gained from a given system of environmental sensing is defined as the optimal Lyapunov exponent with sensing minus the optimal Lyapunov exponent when individuals are unable to sense the environment \cite{donaldson2010fitness,rivoire2011value}. By optimal Lyapunov exponent we mean the maximum Lyapunov exponent achievable over all possible phenotype switching strategies. It is important to find explicit forms for these Lyapunov exponents in order to find the optimal phenotype switching strategies.

Our derivations of Lyapunov exponents hinge upon the assumption that environmental states persist far longer than the timescales of phenotype switching. This assumption, termed the adiabatic limit, justifies a key approximation of phenotype dynamics. This approximation is the following: upon a change in the environment, the distribution of phenotypes in the population reaches a steady state within the new environment before the environment changes again. The validity of the adiabatic approximation rests both on slow environmental dynamics relative to phenotypic dynamics, a reasonable description of many organism-environment pairs. For example, cyanobacteria can alter the pigments they express to allow absorption of different light frequencies through the course of movement through the water column \cite{stomp2008timescale} while amphibious fish can reversibly develop and dispose of lungs with the changes between wet and dry seasons \cite{wright2016amphibious}. In both of these examples, the authors of the corresponding studies emphasize that the separation of environmental and phenotype switching timescales are highly dependent on the specific details of environmental dynamics and organismal physiology and behavior.

Under the stated conditions, we can use the eigenvalues and eigenvectors of each $\mathbf{P}_{\tilde{e}}$ to write the phenotype probability distribution of a population $m$ time steps after the environment has switched from state $e$ to $e'$ when individuals can perfectly sense their environment
\begin{equation}\label{eq:phenotype_distribution_expansion}
\bm{p}_{e\rightarrow{}e'}(m)=\bm{\pi}_{e}+(\bm{\pi}_{e'}-\bm{\pi}_{e}) \exp(-km)
\end{equation}
(To avoid confusion with the environmental state $e$, we use $\exp(\cdot)$ as the exponent of Euler's number.) In the more general case where there is a sensing error probability $\epsilon$, the strategies $\bm{\pi}_{e}$ in Eq. \ref{eq:phenotype_distribution_expansion} are replaced by effective strategies
\begin{equation}
\bm{\rho}_{e}(\epsilon)=(1-\epsilon)\bm{\pi}_{e}+\epsilon\bm{\pi}_{e'}
\end{equation}
so that the phenotype distribution becomes
\begin{equation}\label{eq:phenotype_distribution_expansion_effective}
\bm{p}_{e\rightarrow{}e'}(m,\epsilon)=\bm{\rho}_{e}(\epsilon)+(\bm{\rho}_{e'}(\epsilon)-\bm{\pi}_{e}) \exp(-km).
\end{equation}
This expansion of the phenotype probability distribution with effective strategies allows us to write down an explicit formula for the Lyapunov exponent
\begin{equation}\label{eq:full_lyapunov}
\Lambda^{(a,\epsilon)}_{k}=\sum_{e\in\mathcal{E}}p(e)\log(\mathbf{w}_{e}^{T}\bm{\rho}_{e}(\epsilon))+\frac{1}{2\mu}\sum_{e\in\mathcal{E}}\sum_{m=0}^{\infty}\log\Bigg(1+\frac{\mathbf{w}_{e}^{T}(\bm{\rho}_{e'}(\epsilon)-\bm{\rho}_{e}(\epsilon))}{\mathbf{w}_{e}^{T}\bm{\rho}_{e}(\epsilon)} \exp(-km)\Bigg).
\end{equation}
The superscript $a$ here indicates that Eq. \ref{eq:full_lyapunov} is valid in the adiabatic limit, while $\epsilon$ indicates the error probability in the environmental sensing channel, as before. The subscript $k$ is the rate of phenotype switching. Given our assumptions, a faster phenotype switching rate will always increase the Lyapunov exponent. For this reason, we treat $k$ as a fixed parameter that populations do not optimize over. The inclusion of a realistic cost function for switching speed would allow for non-trivial optimization over $k$, but this is beyond the scope of the present work. In order to determine optimal phenotype switching strategies, we must maximize Eq. \ref{eq:full_lyapunov} with respect to the environment-dependent phenotype distributions $\bm{\pi}_{e_{1}}$ and $\bm{\pi}_{e_{2}}$, which can we changed independently of $k$.

When the error probability is $1/2$, we write the Lyapunov exponent as $\Lambda^{(\eta)}$. Here $\eta$ indicates that the environmental sensing channel is completely uninformative about the environment, while the $a$ superscript and $k$ subscript are removed due to the irrelevance of the adiabatic limit and the phenotype switching rate in this case. The irrelevance of kinetics when $\epsilon=1/2$ is a consequence of assuming that the phenotypes of the population begin at their steady-state frequencies and that the phenotype of interest is not inherited, in contrast to previous work \cite{hufton2018phenotypic}. When there is zero probability of sensing error, we write $\Lambda^{(a,\delta)}_{k}$. Here, $\delta$ indicates a noiseless environmental sensing channel. When the phenotype switching rate is taken to be infinite, we write $k\rightarrow\infty$ in the subscript. The full Lyapunov exponent derivations in different conditions are detailed in Appendix \ref{section:appendixa}. 

\subsection{The fitness value of perfect sensing for $k\rightarrow\infty$ is the mutual information between consecutive environmental states}\label{section:resultsb}

How can a population maximize its long-term growth rate when environmental sensing is unreliable and there is a delay between the time of sensing and the resulting response? More precisely, what phenotype distributions $\hat{\bm{\pi}}_{e_{1}}$ and $\hat{\bm{\pi}}_{e_{2}}$ maximize Eq. \ref{eq:full_lyapunov} given environment switching probabilities, the fitness matrix, the phenotype switching rate, and the sensing error probability? We indicate that a phenotype distribution is optimal with a hat.

We first explore the case where individuals can perfectly sense their environment. The simplest case is when $k\rightarrow{}\infty$, that is, when the phenotype distribution can instantaneously relax to a steady state upon an environmental change, after a unit time step delay. We set the derivations of  $\Lambda^{(a,\delta)}_{k\rightarrow\infty}$ (Eq. \ref{eq:lyapunov_perfect_sensing_large_k}) with respect to $\pi_{\phi_{1},e_{1}}$ and $\pi_{\phi_{1},e_{2}}$ to zero and find
\begin{align}\label{eq:perfect_sensing_k_large_solution1}
\hat{\pi}_{\phi_{1}e_{1},\delta}&=\min\Bigg[\max\Bigg[(1-\alpha)\bigg(1-\frac{w_{\phi_{1}e_{2}}}{w_{\phi_{2}e_{2}}}\bigg)^{-1}+\alpha\bigg(1-\frac{w_{\phi_{1}e_{1}}}{w_{\phi_{2}e_{1}}}\bigg)^{-1},0\Bigg],1\Bigg]\\\label{eq:perfect_sensing_k_large_solution2}
\hat{\pi}_{\phi_{2}e_{1},\delta}&=1-\hat{\pi}_{\phi_{1}e_{1},\delta}\\\label{eq:perfect_sensing_k_large_solution3}
\hat{\pi}_{\phi_{1}e_{2},\delta}&=\min\Bigg[\max\Bigg[\beta\bigg(1-\frac{w_{\phi_{1}e_{2}}}{w_{\phi_{2}e_{2}}}\bigg)^{-1}+(1-\beta)\bigg(1-\frac{w_{\phi_{1}e_{1}}}{w_{\phi_{2}e_{1}}}\bigg)^{-1},0\Bigg],1\Bigg]\\\label{eq:perfect_sensing_k_large_solution4}
\hat{\pi}_{\phi_{2}e_{2},\delta}&=1-\hat{\pi}_{\phi_{1}e_{2},\delta}.
\end{align}
The optimal strategy in this case is closely related to the case where individuals can not sense the environment. When $w_{\phi_{2},e_{1}},w_{\phi_{1},e_{2}}\rightarrow{}0$, the optimal strategy here corresponds to proportional betting on the environmental transition probabilities, rather than on the stationary probabilities of environmental states. Perfect knowledge of the current environmental state allows individuals to condition on the environmental state, rather than relying on an environment-agnostic bet-hedging strategy.

Plugging Eqs. \ref{eq:perfect_sensing_k_large_solution1}-\ref{eq:perfect_sensing_k_large_solution4} into Eq. \ref{eq:lyapunov_perfect_sensing_large_k} yields the optimal Lyapunov exponent
\begin{equation}\label{eq:optimal_lyapunov_perfect}
\hat{\Lambda}^{(a,\delta)}_{k\rightarrow\infty}=\sum_{e\in\mathcal{E}}p(e)\log\bigg(\frac{\det\mathbf{W}}{\Delta{}w_{e'}}\bigg)-H(E_{m+1}|E_{m})
\end{equation}
which is identical to the optimal non-sensing Lyapunov exponent (Eq. \ref{eq:lyapunov_no_sensing}) except that the environmental entropy has been replaced by the entropy of the next environmental state conditioned on the current environmental state. This expression is valid when none of the probabilities in Eqs. \ref{eq:perfect_sensing_k_large_solution1}-\ref{eq:perfect_sensing_k_large_solution4} are equal to 0 or 1. The quantity within the logarithm of Eq. \ref{eq:optimal_lyapunov_perfect} is equivalent to the hypothetical specialist phenotype introduced in \cite{donaldson2010fitness}. When $\hat{\pi}_{\phi_{1}e_{1},\delta}=\hat{\pi}_{\phi_{2}e_{2},\delta}=1$, the optimal Lyapunov exponent is instead
\begin{equation}\label{eq:optimal_lyapunov_perfect_ones}
\hat{\Lambda}^{(a,\delta)}_{k\rightarrow\infty}=\sum_{e_{m}\in\mathcal{E}}\sum_{e_{m+1}\in\mathcal{E}}p(e_{m+1}|e_{m})p(e_{m})\log(w_{e_{m}e_{m+1}})
\end{equation}
where $w_{e_{m}e_{m+1}}$ is the fitness of the phenotype best suited to environment $e_{m}$ when the environment is in state $e_{m+1}$.

The fitness value of information gained through perfect sensing of the current environment (when Eq. \ref{eq:optimal_lyapunov_perfect} is valid) is
\begin{equation}
\hat{\Lambda}^{(a,\delta)}_{k\rightarrow\infty}-\hat{\Lambda}^{(\eta)}=H(E)-H(E_{m+1}|E_{m})=I(E_{m};E_{m+1}),
\end{equation}
the mutual information between the environmental state and the immediately following environmental state (Fig. \ref{fig:value_of_information_finite_k}). If the environment is IID, the mutual information between the current and following environmental states is zero, so that as expected the value of information is zero. This demonstrates that the fitness value of information is bounded from above by $I(E_{m};E_{m+1})$, analogous to the results of Donaldson-Matasci, Bergstrom, and Lachmann \cite{donaldson2010fitness}. In general, the Lyapunov exponent with no information subtracted from the Lyapunov exponent with perfect information is
\begin{align}\label{eq:suboptimal_fvoi}\nonumber
\Lambda^{(a,\delta)}_{k\rightarrow\infty}-\Lambda^{(\eta)}=&\sum_{e_{m}\in\mathcal{E}}\sum_{e_{m+1}\in\mathcal{E}}p(e_{m+1}|e_{m})p(e_{m})\log\bigg(\pi_{e_{m}e_{m+1}}+\frac{w_{e_{m}'e_{m}}}{\Delta{}w_{e_{m}}}\bigg)\\
&-\sum_{e_{m}\in\mathcal{E}}p(e_{m})\log\bigg(\pi_{e_{m}}+\frac{w_{e_{m}'e_{m}}}{\Delta{}w_{e_{m}}}\bigg).
\end{align}
Maximizing Eq. \ref{eq:suboptimal_fvoi} will not yield the fitness value of information, which is the difference between the maximal Lyapunov exponents rather than the maximal difference in the Lyapunov exponents.

\begin{figure}[ht!]
    \centering
    \includegraphics[width=.75\textwidth]{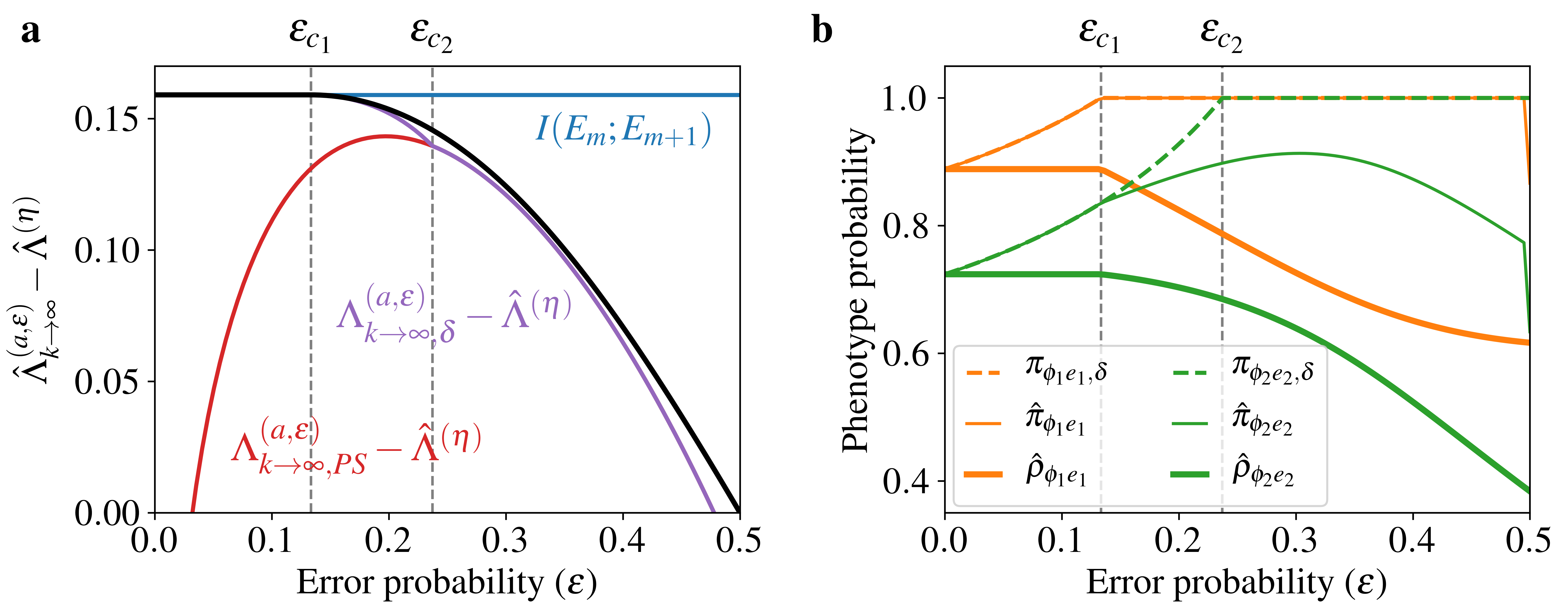}
    \caption{{\bf The effects of environment sensing error probability on long-term growth rate.} (a) The fitness value of information ($\hat{\Lambda}^{(a,\epsilon)}_{k\rightarrow\infty}-\hat{\Lambda}^{(\eta)}$) for $k\rightarrow\infty$ is shown as a black curve as a function of $\epsilon$. This is the value of information achieved by the optimal strategy, as found through numerical optimization. The value of perfect information, $\hat{\Lambda}^{(a,\delta)}_{k\rightarrow\infty}-\hat{\Lambda}^{(\eta)}=I(E_{m}|E_{m+1})$ is shown in blue, representing an upper bound on the value of information. The value of information when a pure strategy is used ($\pi_{\phi_{1}e_{1}}=\pi_{\phi_{2}e_{2}}=1$) is shown in red as a function of $\epsilon$. Finally, the value of information for the effective strategy resembling the optimal strategy when $\epsilon=0$ as closely as possible is shown in purple. This corresponds to the optimal strategy when $\epsilon\leq\epsilon_{c_{1}}$, but is no longer optimal for $\epsilon>\epsilon_{c_{1}}$. (b) Optimal and suboptimal phenotype expression probabilities as a function of $\epsilon$. The dashed lines follow the strategy ($\pi_{\phi_{1}e_{1},\delta}$ and $\pi_{\phi_{2}e_{2},\delta}$) yielding the effective strategy that most closely matches that for when $\epsilon=0$. These probabilities yield the purple curve in (a). The thin solid lines follow the optimal strategy, corresponding to the black curve in (a). The thick solid lines follow the effective strategy resulting from the optimal strategy. Model parameters are fixed at $w_{\phi_{1}e_{1}}=5.0$, $w_{\phi_{2}e_{1}}=0.01$, $w_{\phi_{1}e_{2}}=1.0$, $w_{\phi_{2}e_{2}}=10.0$, $\alpha=0.2$, and $\beta=0.25$. This figure was created in a Python 3.7 Jupyter Notebook \protect\cite{kluyver2016jupyter} with Matplotlib 3.1.1 \protect\cite{hunter2007matplotlib}.}
    \label{fig:value_of_information_infinite_k}
\end{figure}

\subsection{Subject to given conditions, imperfect sensing has no impact on the fitness value of information below a critical error probability}\label{section:resultsc}

Any organism is limited in how accurately it can sense the state of its surroundings. In this section, we explore the consequences of imperfect environmental sensing on the population growth model discussed above. With environmental sensing error probability $\epsilon$, the effective strategy adopted in an environmental state is
\begin{align}\label{eq:eff_strat1}
\rho_{\phi_{1},e_{1}}(\epsilon)&=(1-\epsilon)\pi_{\phi_{1},e_{1}}+\epsilon\pi_{\phi_{1},e_{2}}\\\label{eq:eff_strat2}
\rho_{\phi_{2},e_{1}}(\epsilon)&=1-\rho_{\phi_{1},e_{1}}(\epsilon)\\\label{eq:eff_strat3}
\rho_{\phi_{1},e_{2}}(\epsilon)&=\epsilon\pi_{\phi_{1},e_{1}}+(1-\epsilon)\pi_{\phi_{1},e_{2}}\\\label{eq:eff_strat4}
\rho_{\phi_{2},e_{2}}(\epsilon)&=1-\rho_{\phi_{1},e_{2}}(\epsilon)
\end{align}
so that the strategy adopted when an individual perceives environment $e_{1}$, for example, is described by $\pi_{\phi_{1},e_{1}}$, but the actual strategy adopted in $e_{1}$ is an average of phenotypes expressed when $e_{1}$ is correctly perceived and when $e_{2}$ is mistakenly perceived. 

Unfortunately, it is difficult to explicitly find the general optimal phenotype expression strategy when $0<\epsilon<1/2$. We instead performed computational optimization of the Lyapunov exponent over $\pi_{\phi_{1},e_{1}}$ and $\pi_{\phi_{1},e_{2}}$ using the optimization package within SciPy 1.3.1 \cite{scipy}.

When $\epsilon=0$, the perfect sensing case is recovered and the optimal Lyapunov exponent is equal to $\hat{\Lambda}^{(a,\delta)}_{k\rightarrow\infty}$. On the other hand, when $\epsilon=1/2$ the no-sensing case is recovered, so that the optimal Lyapunov exponent is $\hat{\Lambda}^{(\eta)}$. With $0<\epsilon<1/2$, $\hat{\Lambda}^{(a,\epsilon)}_{k\rightarrow\infty}$ is a non-increasing function of $\epsilon$ (Fig. \ref{fig:value_of_information_infinite_k}a). 

There can be two critical values of $\epsilon$ at which the nature of possible strategies changes. Remarkably, for $\epsilon<\epsilon_{c_{1}}$, there exists a strategy which yields the same long-term growth rate as when $\epsilon=0$, so that $\hat{\Lambda}^{(a,\epsilon)}_{k\rightarrow\infty}=\hat{\Lambda}^{(a,\delta)}_{k\rightarrow\infty}$. For $\epsilon_{c_{1}}<\epsilon<\epsilon_{c_{2}}$, it is only possible to adopt the same optimal effective strategy as in the perfect sensing case in one of the two environmental states, while when the other environmental state is perceived the corresponding better-adapted phenotype is expressed with a probability of 1 (Fig. \ref{fig:value_of_information_infinite_k}). 

The values of $\epsilon_{c_{1}}$ and $\epsilon_{c_{2}}$ in terms of the optimal strategy for $\Lambda^{(a,\delta)}_{k\rightarrow\infty}$ (Eqs. \ref{eq:perfect_sensing_k_large_solution1}-\ref{eq:perfect_sensing_k_large_solution4}) are
\begin{align}
\epsilon_{c_{1}}&=\min\bigg(\frac{\hat{\pi}_{\phi_{1}e_{2},\delta}}{\hat{\pi}_{\phi_{1}e_{2},\delta}+\hat{\pi}_{\phi_{1}e_{1},\delta}},\frac{\hat{\pi}_{\phi_{2}e_{1},\delta}}{\hat{\pi}_{\phi_{2}e_{1},\delta}+\hat{\pi}_{\phi_{2}e_{2},\delta}}\bigg)\\
\epsilon_{c_{2}}&=\max\bigg(\frac{\hat{\pi}_{\phi_{1}e_{2},\delta}}{\hat{\pi}_{\phi_{1}e_{2},\delta}+\hat{\pi}_{\phi_{1}e_{1},\delta}},\frac{\hat{\pi}_{\phi_{2}e_{1},\delta}}{\hat{\pi}_{\phi_{2}e_{1},\delta}+\hat{\pi}_{\phi_{2}e_{2},\delta}}\bigg)
\end{align}
which is a result of finding the smallest value of $\epsilon$ where $\hat{\pi}_{\phi_{1}e_{1}}=1$ and the smallest value of $\epsilon$ where $\hat{\pi}_{\phi_{1}e_{2}}=0$. The quantity $\frac{\hat{\pi}_{\phi_{1}e_{2},\delta}}{\hat{\pi}_{\phi_{1}e_{2},\delta}+\hat{\pi}_{\phi_{1}e_{1},\delta}}$ is the probability that environment is in state $e_{2}$ given that the phenotype is in state $\phi_{1}$ while  $\frac{\hat{\pi}_{\phi_{2}e_{1},\delta}}{\hat{\pi}_{\phi_{2}e_{1},\delta}+\hat{\pi}_{\phi_{2}e_{2},\delta}}$ is the probability that environment is in state $e_{1}$ given that the phenotype is in state $\phi_{2}$, when the optimal strategy for the perfect sensing case is implemented. These probabilities can be interpreted as the average probability of error of an individual's internal predictions of the environmental state when the optimal strategy for the perfect sensing case is adopted. As mentioned before, below $\epsilon_{c_{1}}$ it is possible to adopt $\pi_{\phi_{1}e_{1}}$ and $\pi_{\phi_{1}e_{2}}$ such that the effective strategies match the optimal strategy for $\epsilon=0$. This strategy is
\begin{align}\label{eq:imperfect_sensing_k_large_solution1}
\hat{\pi}_{\phi_{1}e_{1}}&=\min\Bigg[\max\Bigg[\Bigg(\frac{(1-\epsilon)(1-\alpha)-\epsilon\beta}{1-w_{\phi_{1}e_{2}}/w_{\phi_{2}e_{2}}}+\frac{(1-\epsilon)\alpha-\epsilon(1-\beta)}{1-w_{\phi_{1}e_{1}}/w_{\phi_{2}e_{1}}}\Bigg)(1-2\epsilon)^{-1},0\Bigg],1\Bigg]\\\label{eq:imperfect_sensing_k_large_solution2}
\hat{\pi}_{\phi_{2}e_{1}}&=1-\hat{\pi}_{\phi_{1}e_{1}}\\\label{eq:imperfect_sensing_k_large_solution3}
\hat{\pi}_{\phi_{1}e_{2}}&=\min\Bigg[\max\Bigg[\Bigg(\frac{(1-\epsilon)\beta-\epsilon(1-\alpha)}{1-w_{\phi_{1}e_{2}}/w_{\phi_{2}e_{2}}}+\frac{(1-\epsilon)(1-\beta)-\epsilon\alpha}{1-w_{\phi_{1}e_{1}}/w_{\phi_{2}e_{1}}}\Bigg)(1-2\epsilon)^{-1},0\Bigg],1\Bigg]\\\label{eq:imperfect_sensing_k_large_solution4}
\hat{\pi}_{\phi_{2}e_{2}}&=1-\hat{\pi}_{\phi_{1}e_{2}}.
\end{align}
When $\epsilon>\epsilon_{c_{1}}$, this strategy is no longer optimal and corresponds to $\pi_{\phi_{1}e_{1},\delta}$ and $\pi_{\phi_{2}e_{2},\delta}$ in Fig. \ref{fig:value_of_information_infinite_k}b. Between $\epsilon_{c_{1}}$ and $\epsilon_{c_{2}}$ the effective strategy can be made to match the optimal strategy with perfect information for only one environmental state, although it is not necessarily optimal to do so (Fig. \ref{fig:value_of_information_infinite_k}). We emphasize that $\epsilon_{c_{2}}$ is more of technical interest than of any biological importance, in constrast to $\epsilon_{c_{1}}$ which has significant biological implications.  

\begin{figure}[ht!]
    \centering
    \includegraphics[width=.75\textwidth]{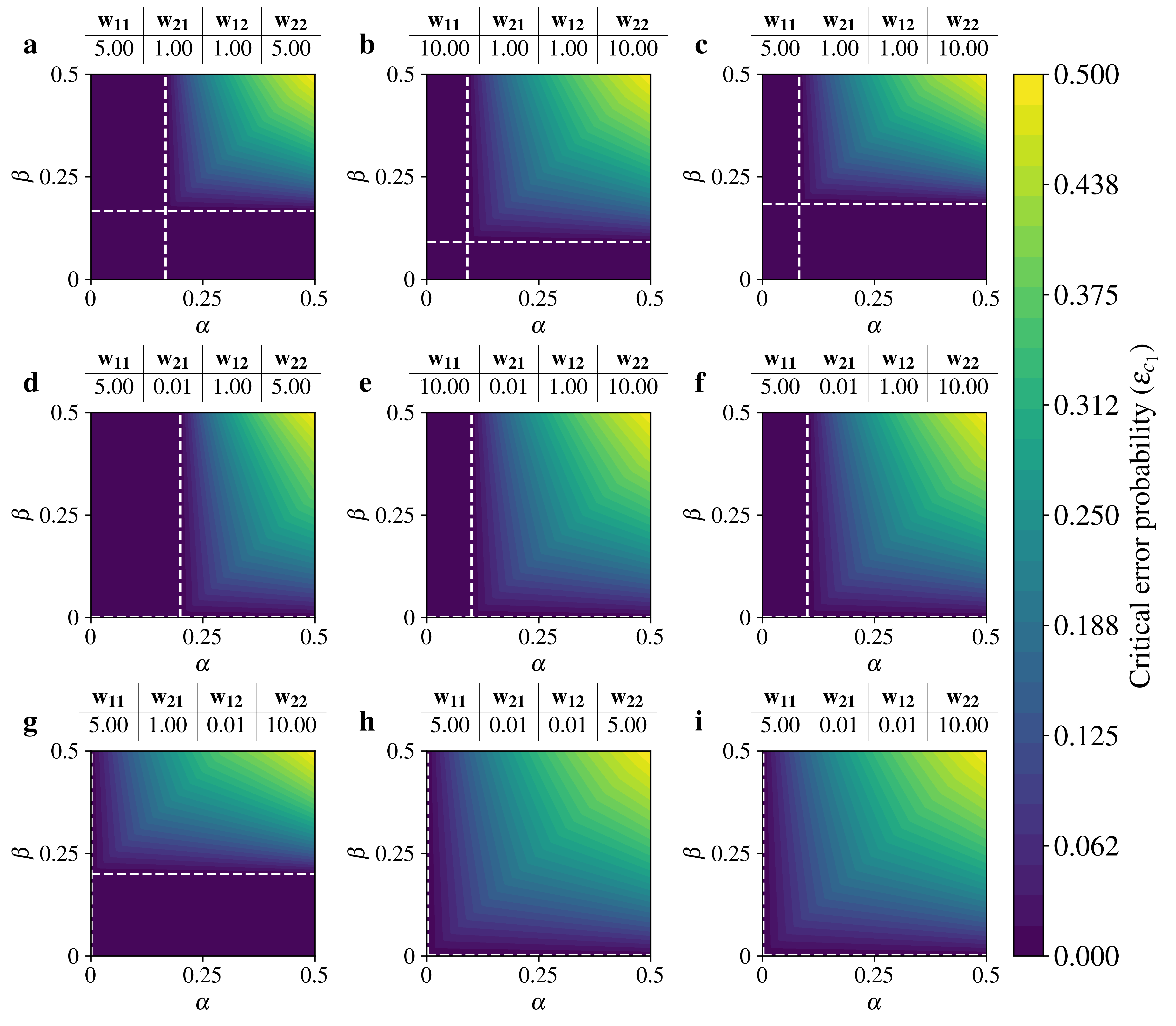}
    \caption{{\bf The critical error probability in different conditions.} The value of the critical error probability $\epsilon_{c_{1}}$ in statistically different environments and with different fitness values when $k\rightarrow\infty$. Dotted white lines indicate the values of $\alpha$ and $\beta$ above which $\epsilon_{c_{1}}$ is greater than zero. This figure was created in a Python 3.7 Jupyter Notebook \protect\cite{kluyver2016jupyter} with Matplotlib 3.1.1 \protect\cite{hunter2007matplotlib}.}
    \label{fig:critical_error_probability}
\end{figure}

When is $\epsilon_{c_{1}}$ greater than zero? In other words, when is it possible to achieve $\Lambda^{(a,\epsilon)}_{k\rightarrow\infty}=\Lambda^{(a,\delta)}_{k\rightarrow\infty}$ for $\epsilon>0$? Assuming fixed fitness values, the environmental transition probabilities must satisfy both
\begin{align}\label{eq:crit_cond1}
\alpha&>\frac{w_{\phi_{1}e_{2}}\Delta{}w_{e_{1}}}{\det\mathbf{W}}\\\label{eq:crit_cond2}
\beta&>\frac{w_{\phi_{2}e_{1}}\Delta{}w_{e_{2}}}{\det\mathbf{W}}
\end{align}
in order for $\epsilon_{c_{1}}$ to be greater than zero. This indicates that as the environmental transition probabilities increase, so can $\epsilon_{c_{1}}$ once the conditions in Eqs. \ref{eq:crit_cond1}-\ref{eq:crit_cond2} are met. The critical error probability $\epsilon_{c_{1}}$ does indeed increase once these conditions are met, as shown in Fig. \ref{fig:critical_error_probability}, where Fig. \ref{fig:critical_error_probability}f uses the fitness values used in all other figures. We can see that in the cases analyzed, a critical error probability of 1/2 is only possible when $\alpha=\beta=1/2$, corresponding to the case when no information about the following environmental state is available (Fig. \ref{fig:critical_error_probability}).

The conditions for a positive $\epsilon_{c_{1}}$ in Eqs. \ref{eq:crit_cond1}-\ref{eq:crit_cond2} can be interpreted as follows. The determinant of $\mathbf{W}$ in the denominators is, in some sense, a measure of the degree of phenotype specialization. If both phenotypes have similar fitnesses in both environments, representing generalist phenotypes, $\det\mathbf{W}$ will be small so that the environment must rapidly switch for the critical error probability to be greater than zero. The numerators are the fitness of the maladapted fitness in the destination environment multiplied by the fitness advantage of the better-adapted phenotype over the maladapted phenotype in the source environment ($\Delta{}w_{e}$). The more diagonal the fitness matrix is, the smaller the lower bound will be on environment switching probabilities. In biological terms, greater specialization of phenotypes allows for optimal performance of noisy sensors in more predictable, slowly fluctuating environments. In an extreme case with very small off diagonal fitnesses ($w_{\phi_{1}e_{1}}>>w_{\phi_{2}e_{1}}$ and $w_{\phi_{2}e_{2}}>>w_{\phi_{1}e_{2}}$), the bounds are approximately
\begin{align}
\alpha&>\frac{w_{\phi_{1}e_{2}}}{w_{\phi_{2}e_{2}}}\\
\beta&>\frac{w_{\phi_{2}e_{1}}}{w_{\phi_{1}e_{1}}}.
\end{align}
In this case, $\epsilon_{c_{1}}$ exists for nearly any pair of environmental switching probabilities. This suggests that populations with specialist phenotypes are able to better endure noisy sensing mechanisms in wide ranges of environmental conditions. Alternatively, if the lower bounds on the environment switching probabilities are high as with generalist phenotypes, a noisy sensor can only be optimal in very unpredictable environments, if at all.

\begin{figure}[ht!]
    \centering
    \includegraphics[width=.75\textwidth]{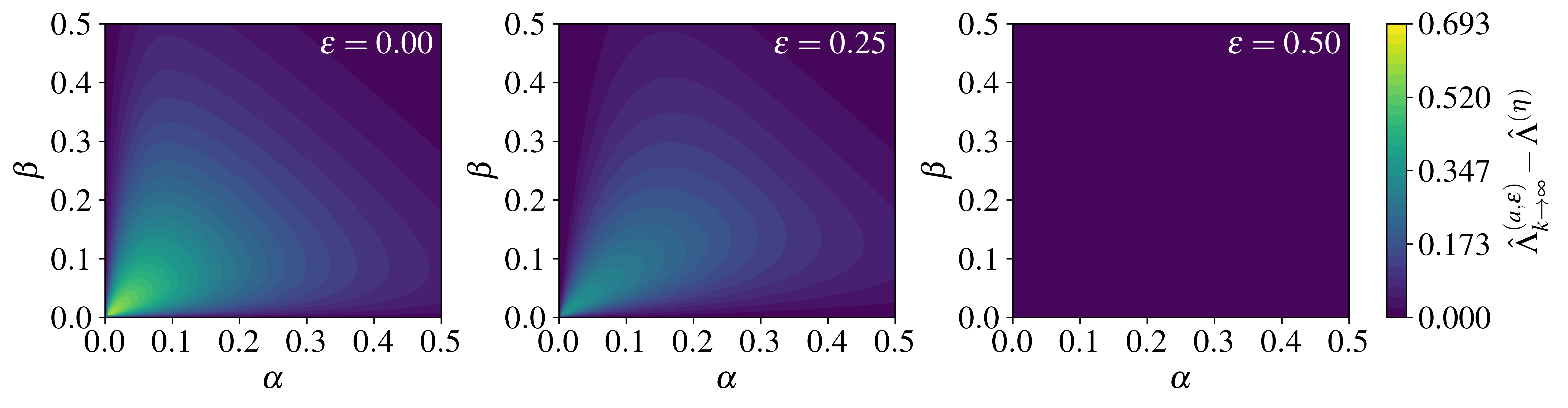}
    \caption{{\bf The value of information in statistically different environments with different sensing error probabilities.} As in the text, $\epsilon$ is the environmental sensing error probability and $\alpha$ and $\beta$ are environmental transition probabilities. Fitness values are fixed at $w_{\phi_{1}e_{1}}=5.0$, $w_{\phi_{2}e_{1}}=0.01$, $w_{\phi_{1}e_{2}}=1.0$, $w_{\phi_{2}e_{2}}=10.0$. This figure was created in a Python 3.7 Jupyter Notebook \protect\cite{kluyver2016jupyter} with Matplotlib 3.1.1 \protect\cite{hunter2007matplotlib}.}
    \label{fig:value_of_information_different_environments}
\end{figure}

We also examine the value of information in different environmental conditions (Fig. \ref{fig:value_of_information_different_environments}). The value of information is maximal as both $\alpha$ and $\beta$ go to zero, as long as $\epsilon<1/2$. In the case where $\epsilon=0$, the fitness value of information is maximal for any $\alpha$, $\beta$ pair while when $\epsilon=1/2$ the fitness value of information is zero regardless of the environmental transition probabilities. 

This makes intuitive sense, as very small environmental transition probabilities mean that conditioning on the current environmental state provides a very accurate prediction of the subsequent environmental state. This highlights an apparent trade-off between the value of sensing the environment and the robustness of sensing mechanism in terms of error probability, which depends on the statistical nature of the environment. In very slowly varying environments, the fitness value of information is large but even small error probabilities will reduce it as a result of exceeding a small $\epsilon_{c_{1}}$. In more rapidly varying environment, the fitness value of information is uniformly smaller by comparison, but sensing errors will have no effect on the Lyapunov exponent below a larger $\epsilon_{c_{1}}$ value. 

\subsection{Slower phenotype switching rates reduce the fitness value of information while increasing robustness to imperfect sensing}\label{section:resultsd}

How does a finite rate of phenotype switching constrain the fitness value of information? What strategies are optimal when phenotypes and the environment switch at similar rates? Again, we can not address these questions by explicitly finding strategies that maximize the Lyapunov exponent. Instead, we investigate the case of finite phenotype switching rates using numerical optimization.

\begin{figure}[ht!]
    \centering
    \includegraphics[width=.75\textwidth]{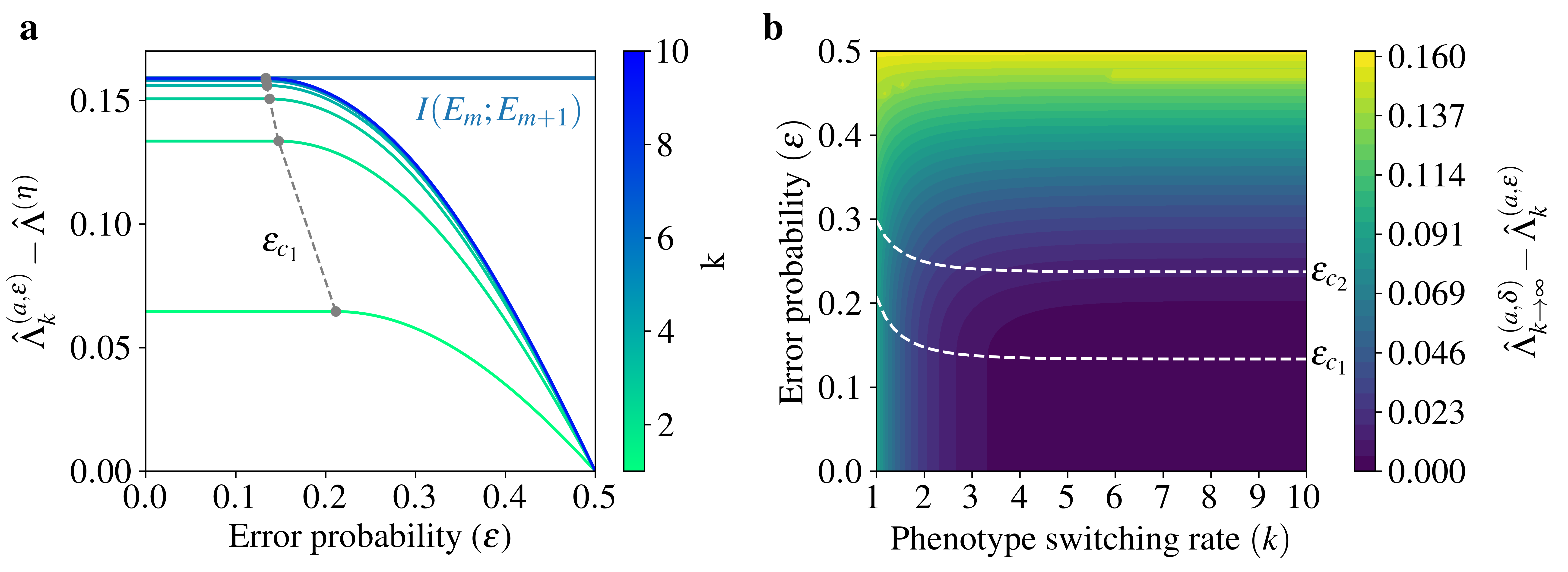}
    \caption{{\bf The fitness value and cost of imperfect information with finite phenotype switching rate.} The effects of phenotype switching rate and environment sensing error probability on the fitness value of information and fitness cost of imperfect sensing. Model parameters are fixed at $w_{\phi_{1}e_{1}}=5.0$, $w_{\phi_{2}e_{1}}=0.01$, $w_{\phi_{1}e_{2}}=1.0$, $w_{\phi_{2}e_{2}}=10.0$, $\alpha=0.2$, and $\beta=0.25$. (a) The fitness value of information, $\hat{\Lambda}^{(a,\epsilon)}_{k\rightarrow\infty}-\hat{\Lambda}^{(\eta)}$ is shown as a function of $\epsilon$ at several values of the phenotype switching rate $k$, denoted by the colorbar. The fitness value of information with perfect environmental sensing and infinite phenotype switching rate, $I(E_{m};E_{m+1})$, is shown in light blue. As $k$ decreases, the interval of $\epsilon$ values for which the fitness value is the same as if $\epsilon=0$ increases. (b) The fitness cost of imperfect information and finite phenotype switching rates. There is a large region below $\epsilon\approx{}0.2$ and above $k\approx{}3.5$ where the fitness cost is close to zero and largely insensitive to changes in either $\epsilon$ or $k$. Outside this region of parameter space, the fitness cost increases far more rapidly as a function of $k$ than of $\epsilon$. Note the dependence of $\epsilon_{c_{1}}$ and $\epsilon_{c_{2}}$ on $k$. This figure was created in a Python 3.7 Jupyter Notebook \protect\cite{kluyver2016jupyter} with Matplotlib 3.1.1 \protect\cite{hunter2007matplotlib}.}
    \label{fig:value_of_information_finite_k}
\end{figure}

As expected, when $k$ is large enough $\Lambda^{(a,\epsilon)}_{k}$ changes with $\epsilon$ much as $\Lambda^{(a,\epsilon)}_{k\rightarrow\infty}$ does (Fig. \ref{fig:value_of_information_finite_k}a). With smaller values of $k$, the value of information decreases for all error probabilities, because the population takes longer to adjust to environmental changes. We note that as $k$ nears $-\log(1-\alpha-\beta)$ the adiabatic assumption no longer applies because the rate of phenotype switching approaches the rate of environment switching. Interestingly, the critical error probabilities $\epsilon_{c_{1}}$ and $\epsilon_{c_{2}}$ increase as $k$ decreases. This is because the slower the population phenotype frequencies equilibrate upon a change in the environmental state, the more advantageous sensing errors in the preceding environmental state become during the relaxation period.

As mentioned earlier, our model does not account for any metabolic costs accrued by increasing the rate of phenotype switching rate nor does it account for biophysical limits to the switching rate. Although an in-depth exploration of the metabolic costs to maintaining a certain rate of phenotype switching would require careful analysis of a particular biological system, we posit that at the very least metabolic cost should be a strictly increasing function of the switching rate, and the resulting fitness cost should be a strictly increasing function of the metabolic cost. If this holds true, the fitness value of information will first increase with $k$ before dropping due to large values of the direct switching cost. This allows for the case where the optimal value of $k$ is not the same for all values of $\epsilon$, so that a smaller phenotype switching rate is more favorable at large values of $\epsilon$ than at small $\epsilon$ values due to the inverse relationship between $\epsilon_{c_{1}}$ and $k$. 

\section{Discussion}

We have explored the fitness value of information in a simple model of Markov phenotype switching in a Markov environment. Consistent with previous results, we found that the fitness value of information can be at most the mutual information between the environmental state at one time and the immediately following environmental state. Intriguingly, we found that if certain environmental conditions are met (Eqs. \ref{eq:crit_cond1}-\ref{eq:crit_cond2}), there is a critical error probability in environmental sensing below which a population can achieve the same fitness as if the error probability were zero. 

The existence of this critical error probability has several intriguing biological implications. First, a population of organisms with imperfect sensory systems can achieve an optimal long-term growth rate. We emphasize that our model does not account for any time integration or collective information sharing in the population, either of which could improve sensory performance, suggesting that an optimal strategy may be possible without any complicated information processing systems. Given the energetic costs of accurate sensing, one might expect evolution of sensory systems towards the critical error probability, which would reduce the metabolic costs of sensing as much as possible while still allowing for maximal population growth rate. However, we emphasize that our model deliberately ignores a number of potentially relevant factors and these predictions require further theoretical and experimental inquiry. 

Second, in the case where phenotype switching rate is sufficiently large, there is a trade-off between the fitness value of information and the robustness of the sensing mechanism to error, depending on the statistical nature of the environment. From our results we expect that predictable environments, which allow populations to capitalize on sensory information, also lead to optimal long-term growth rates that are more sensitive to sensory error probability, in that $\epsilon_{c_{1}}$ is generally small. Alternatively, less predictable environments lead to smaller optimal long-term growth rates that are more tolerant of noisy sensory mechanisms, in that $\epsilon_{c_{1}}$ can generally be large. This could mean that less predictable environments allow for larger diversity in sensory mechanisms while in more predictable environments a lower critical error probability leads to selection against a greater number of sensory mechanisms. Additionally, the conditions for a non-zero $\epsilon_{c_{1}}$ (Eqs. \ref{eq:crit_cond1}-\ref{eq:crit_cond2}) become less strict for more specialized phenotypes, indicating that specialization allows for implementation of optimal strategies with noisy sensors in more predictable environments.

We have not explicitly included fitness costs associated with environmental sensing in our model. This is a clear limitation, as accurate sensing will accrue costs, metabolic and otherwise, that counteract the fitness benefits of environmental awareness. Recent work has advanced understanding of the biophysical relationships between energy consumption and signaling accuracy at the level of cellular signaling \cite{mehta2012energetic,lang2014thermodynamics,sartori2014thermodynamic,mehta2016landauer}, suggesting fundamental trade-offs between costs and signaling fidelity. These findings support our very general assertion of a monotonically increasing fitness cost with decreasing error probability. If this relationship between signaling and cost holds, and the costs are significant enough for selection to act upon, it is clear within our model that when $\epsilon_{c_{1}}>0$ a noisy sensor near $\epsilon_{c_{1}}$ will yield a higher optimal Lyapunov exponent for a population than a perfect sensor will. Though we do not expect the details beyond the stated monotonic relationship to alter this outcome, an intriguing avenue for future work could be to incorporate findings on biophysical cost-accuracy trade-offs into our model.

Other extensions to our model could include density-dependent fitness, finite populations size, noisy global cues, spatial fluctuations, finite time horizons, and non-stationary environments. However, treatment of some of these factors may be better suited to different types of models, such as birth-death models in the case of finite population size and density-dependent fitness \cite{ashcroft2014fixation}. Birth-death models allow for calculation of fixation probabilities and times, which should be more useful for defining evolutionary success if there is a carrying capacity as the long-term growth rate is always zero in this case. A time-dependent environmental Markov chain could potentially favor bet-hedging among sensing mechanisms, which we have not considered in the current study. Additionally, it could be of interest to investigate environments following higher-order Markov processes, so that memory of previous environmental states could be advantageous. As there should be a metabolic cost to memory, it could also be interesting to explore how the costs and benefits of memory can be optimally balanced.

A possible experimental test of our results could involve a bacterial strain engineered to attain varying expression levels of a receptor, which controls response to some external factor that can be manipulated by the experimenters. The environmental factor could be made to effectively follow a two-state Markov process, and the growth rate of bacterial populations with different levels of receptor expression could be measured in a chemostat. The bacteria could be grown in these conditions for a number of generations in order to promote adoption of an optimal switching strategy. Then measuring the growth rate of the population should yield an experimental growth rate $\hat{\Lambda}^{(a,x)}_{\text{experimental}}$ as a function of receptor expression, $x$, which includes the costs of sensing. A control experiment would measure the growth of the same populations in the same environment but where the receptors have a mutation rendering them non-functional. The intent of this control is to measure the cost of increasing sensory abilities without any adaptive value, where the measured growth rate is $\hat{\Lambda}^{(a,x)}_{\text{control}}$. In order to obtain a growth cost $\hat{\Lambda}^{(a,x)}_{\text{cost}}$ as a function of sensory ability from the control experiment, the control growth rate at each $x$ must be subtracted from growth rate at the lowest measured receptor expression rate $x_{min}$
\begin{equation}
\hat{\Lambda}^{(a,x)}_{\text{cost}}=\hat{\Lambda}^{(a,x_{min})}_{\text{control}}-\hat{\Lambda}^{(a,x)}_{\text{control}}.
\end{equation}
To obtain the purely adaptive value of information, one could then subtract the cost from the experimental growth rate
\begin{equation}
\hat{\Lambda}^{(a,x)}_{\text{adaptive}}=\hat{\Lambda}^{(a,x)}_{\text{experimental}}-\hat{\Lambda}^{(a,x)}_{\text{cost}}
\end{equation}

Our model predicts that under the right circumstances, the greatest value of $\hat{\Lambda}^{(a,x)}_{\text{adaptive}}$ should be observed at an intermediate receptor expression level. It may be difficult to distinguish a genuine analogue to a critical error probability from simple diminishing returns in the value of increasing receptor expression. However, statistical testing could be useful to distinguish between the two cases. A similar experimental setup to Acar, Mettetal, and Van Oudenaarden \cite{acar2008stochastic} may be able to accomplish these experimental goals.

\begin{acknowledgments}
This work was supported by the U.S. Defense Advanced Research Projects Agency RadioBio program under grant number HR001117C0125.
\end{acknowledgments}

\appendix
\section{Derivation of the Lyapunov exponents}\label{section:appendixa}

Using a similar approach to Kussell and Leibler \cite{kussell2005phenotypic}, we first derive the Lyapunov exponent in the case of a slowly varying environment for a population of individuals with two phenotypic states and with the ability to perfectly sense the environmental state. As stated before, we assume that phenotypes are not heritable. Rewriting Eq. \ref{eq:growth}, the total population size at time step $M$ is,
\begin{equation}
N_{M}=\exp\Bigg(\sum_{m=0}^{M-1}\log{\overline{w}\big(m,e(m)\big)}\Bigg)N_{0}
\end{equation}
where $\overline{w}\big(m,e(m)\big)$ is the time-dependent mean fitness of the population at time step $m$. Rearranging and using the definition of the dominant Lyapunov exponent \cite{metz1992should,rivoire2011value}, we find
\begin{equation}
\Lambda\equiv\lim_{M\rightarrow\infty}\frac{1}{M}\log\Bigg(\frac{N_{M}}{N_{0}}\Bigg)=\lim_{M\rightarrow\infty}\frac{1}{M}\sum_{m=0}^{M-1}\log{\overline{w}\big(m,e(m)\big)}.
\end{equation}
We then break $M$ down into $L$ time periods where the environment is constant, rewriting $M=M_{L}=\sum_{l=1}^{L}M_{l}$, where $M_{L}=0$. We can then rewrite $\Lambda$ as
\begin{equation}
\Lambda=\lim_{L\rightarrow\infty}\frac{1}{M_{L}}\sum_{l=1}^{L}\sum_{m=M_{l-1}}^{M_{l}-1}\log{\overline{w}\big(m,e(m)\big)}.
\end{equation}
Next, we make our main assumption that each $M_{l}$ is long enough for the population to reach an equilibrium distribution of phenotypes. This is the so-called adiabatic limit \cite{kussell2005phenotypic,rivoire2011value}. We now have
\begin{align}\nonumber
\Lambda^{(a)}&=\lim_{L\rightarrow\infty}\frac{1}{M_{L}}\sum_{e\in\mathcal{E}}\sum_{l=1}^{L^{e}}\sum_{m=0}^{M_{l}^{e}-1}\log{\overline{w}\big(m,e\big)}\\\label{eq:adiabatic}
&=\lim_{L\rightarrow\infty}\frac{1}{L\mu}\sum_{e\in\mathcal{E}}L^{e}\frac{1}{L^{e}}\sum_{l=1}^{L^{e}}\sum_{m=0}^{M_{l}^{e}-1}\log{\overline{w}\big(m,e\big)}
\end{align}
where $\Lambda^{(a)}$ is the Lyapunov exponent in the adiabatic limit and the environment in each interval is no longer time-dependent. $M_{l}^{e}$ is the $l^{th}$ time period spent in environment $e$, $L^{e}$ is the number of times that the environment transitioned to $e$ from the other possible environmental state, and $\mu=\lim_{L\rightarrow\infty}\frac{M_{L}}{L}$ is the average time (in number of time steps) spent in any environmental state. Because there are only two environmental states, $\lim_{L\rightarrow\infty}L^{e}/L=1/2$, and using the law of large numbers
\begin{equation}\label{eq:lln}
\Lambda^{(a)}=\frac{1}{2\mu}\sum_{e\in\mathcal{E}}\Bigg\langle\sum_{m=0}^{M_{l}^{e}-1}\log{\overline{w}\big(m,e\big)}\Bigg\rangle_{M_{l}^{e}}
\end{equation}
where $\langle\cdot\rangle_{M_{l}^{e}}$ denotes the mean with respect to time periods in environment $e$. As a consequence of assuming that phenotypes are not inherited, the mean fraction of the population in each phenotype does not depend on the fitness of the phenotype. Instead, we can use the eigenvalues and eigenvectors of the phenotypic Markov chains, which are easy to calculate for our two-state model. We recall the definition of $\bm{\pi}_{\tilde{e}}$ as the normalized right (column) eigenvector of $\mathbf{P}_{\tilde{e}}$ corresponding to an eigenvalue of $\lambda_{\tilde{e}}=1$, which describes the steady-state distribution of phenotypes in perceived environmental state $\tilde{e}$. We also recall $\mathbf{l}_{\tilde{e}}$ and $\mathbf{v}_{\tilde{e}}$, the left (row) and right (column) eigenvectors of $\mathbf{P}_{\tilde{e}}$ corresponding to the eigenvalue $\lambda_{\tilde{e}}=1-\chi_{\tilde{e}}-\omega_{\tilde{e}}$. Immediately after the environment has switched from state $e'$ to $e$, if individuals can perfectly sense it, the mean fitness is $\mathbf{w}_{e}^{T}\bm{\pi}_{e'}$ where $\mathbf{w}_{e}$ is the (column) vector containing the fitness of each phenotype in environment $e$. Over the period of time the environment then spends in state $e$ and if there are only two phenotypes in the population, the mean fitness can then be described as
\begin{equation}\label{eq:time-dependent_growth_rate}
\overline{w}\big(m,e\big)=\mathbf{w}_{e}^{T}\bm{\pi}_{e}+\mathbf{w}_{e}^{T}(\mathbf{l}_{e}\bm{\pi}_{e'})\mathbf{v}_{e}\exp(-km).
\end{equation}
Plugging this into the sum within the average in Eq. \ref{eq:lln}, we have
\begin{align}
\sum_{m=0}^{M_{l}^{e}-1}\log{\overline{w}\big(m,e\big)}&=\sum_{m=0}^{M_{l}^{e}-1}\log\Big(\mathbf{w}_{e}^{T}\bm{\pi}_{e}+\mathbf{w}_{e}^{T}(\mathbf{l}_{e}\bm{\pi}_{e'})\mathbf{v}_{e}\exp(-km)\Big)\\\label{eq:eig_expansion}
&=\sum_{m=0}^{M_{l}^{e}-1}\log(\mathbf{w}_{e}^{T}\bm{\pi}_{e})+\sum_{m=0}^{M_{l}^{e}-1}\log\Bigg(1+\frac{\mathbf{w}_{e}^{T}(\mathbf{l}_{e}\bm{\pi}_{e'})\mathbf{v}_{e}}{\mathbf{w}_{e}^{T}\bm{\pi}_{e}}\exp(-km)\Bigg).
\end{align}
Since we have assumed that each time period $M_{l}^{e}$ is long enough for the distribution of phenotypes within the population to equilibrate, we approximate Eq. \ref{eq:eig_expansion} as
\begin{equation}\label{eq:adiabatic_approx}
\sum_{m=0}^{M_{l}^{e}-1}\log{\overline{w}\big(m,e\big)}\approx\sum_{m=0}^{M_{l}^{e}-1}\log(\mathbf{w}_{e}^{T}\bm{\pi}_{e})+\sum_{m=0}^{\infty}\log\Bigg(1+\frac{\mathbf{w}_{e}^{T}(\mathbf{l}_{e}\bm{\pi}_{e'})\mathbf{v}_{e}}{\mathbf{w}_{e}^{T}\bm{\pi}_{e}}\exp(-km)\Bigg).
\end{equation}
The first term in the sum is easy to evaluate as $\log(\mathbf{w}_{e}^{T}\bm{\pi}_{e})$ is independent of time when the environment is constant. The second term represents the fitness cost of the unit time delay between the actual environmental state and the internal perceived environmental state of individual organisms together with having a finite phenotype switching rate. Both of these factors contribute to a relaxation period immediately following a change in environmental state, in which the population is in the process of changing from the phenotype distribution adapted to the previous environmental state. We refer to this fitness cost as the relaxation cost. Defining $a_{e}\equiv\frac{\mathbf{w}_{e}^{T}(\mathbf{l}_{e}\bm{\pi}_{e'})\mathbf{v}_{e}}{\mathbf{w}_{e}^{T}\bm{\pi}_{e}}$ for the sake of notational compactness and using Eq. \ref{eq:lln}, we write the total relaxation cost as
\begin{equation}\label{eq:relax_cost}
C_{r}=\frac{1}{2\mu}\sum_{e\in\mathcal{E}}\sum_{m=0}^{\infty}\log(1+a_{e}\exp(-km)).
\end{equation}
The biological interpretation of $a_{e}$ becomes clearer if written instead as
\begin{equation}\label{eq:a_interpretation}
a_{e}=\frac{\mathbf{w}_{e}^{T}(\mathbf{l}_{e}\bm{\pi}_{e'})\mathbf{v}_{e}}{\mathbf{w}_{e}^{T}\bm{\pi}_{e}}=\frac{\mathbf{w}_{e}^{T}\bm{\pi}_{e'}-\mathbf{w}_{e}^{T}\bm{\pi}_{e}}{\mathbf{w}_{e}^{T}\bm{\pi}_{e}},
\end{equation}
the relative loss in mean fitness of the population when adapted to the incorrect environment. Writing $a_{e}$ in this manner illuminates a superficial similarity between $a_{e}$ and a selection coefficent, although in this case the two competing ``genotypes'' are the same population when adapted and maladapted to the environment and $a_{e}$ is written in terms of mean population fitness rather than individual fitness. 

We note that for a two-state Markov chain, the stationary distribution $\bm{\pi}_{\tilde{e}}$ and the non-stationary eigenvalue can be changed independently. Consequently, with no fitness cost associated with the speed of phenotype switching, the switching rate in this model can go to infinity without any negative consequences for growth, so that the relaxation cost approaches $\frac{1}{2\mu}\sum_{e\in\mathcal{E}}\log(1+a_{e})$ as $k\rightarrow\infty$. Thus, when $k$ is large, the Lyapunov exponent can be approximated by the Lyapunov exponent obtained by taking $k\rightarrow\infty$
\begin{equation}\label{eq:lyapunov_perfect_sensing_large_k}
\Lambda^{(a,\delta)}_{k\rightarrow\infty}=\sum_{e\in\mathcal{E}}p(e)\log(\mathbf{w}_{e}^{T}\bm{\pi}_{e})+\frac{1}{2\mu}\sum_{e\in\mathcal{E}}\log(1+a_{e})
\end{equation}
where we denote the noiseless information channel as $\delta$ and we have used the fact that $p(e)=\big\langle{}M_{l}^{e}\big\rangle/(2\mu)$. In reality, metabolic costs and biophysical constraints will prevent $k$ from becoming too large. However, this limiting case is most convenient for mathematical analysis and represents the case where $k$ is large enough that the fitness cost due to relaxation periods beyond the unit time delay can be ignored. 

If the phenotype switching rate is not large enough to be approximated as infinite, we can write the Lyapunov exponent as
\begin{equation}\label{eq:lyapunov_expansion}
\Lambda^{(a,\delta)}_{k}=\sum_{e\in\mathcal{E}}p(e)\log(\mathbf{w}_{e}^{T}\bm{\pi}_{e})+\frac{1}{2\mu}\sum_{e\in\mathcal{E}}\Bigg(\log(1+a_{e})+\sum_{n=1}^{\infty}\frac{(-1)^{n+1}a_{e}^{n}}{n(\exp(nk)-1)}\Bigg)
\end{equation}
by expanding $C_{r}$ (Eq. \ref{eq:relax_cost}) in $a_{e_{1}}\exp(-km)$ and $a_{e_{2}}\exp(-km)$. If $k$ is sufficiently large but still finite, we can truncate the infinite sum in Eq. \ref{eq:lyapunov_expansion}, in order to approximate $\Lambda^{(a,\delta)}_{k}$. Throughout this work, we use a second-order approximation
\begin{equation}\label{eq:lyapunov_approx}
\Lambda^{(a,\delta)}_{k}\approx\sum_{e\in\mathcal{E}}p(e)\log(\mathbf{w}_{e}^{T}\bm{\pi}_{e})+\frac{1}{2\mu}\sum_{e\in\mathcal{E}}\Bigg(\log(1+a_{e})+\frac{a_{e}}{\exp(k)-1}-\frac{a_{e}^{2}}{2(\exp(2k)-1)}\Bigg).
\end{equation}

In the more general case where there is a probability $\epsilon$ of error in perceiving the environmental state, the phenotype distribution $\bm{\pi}_{e}$ in Eq. \ref{eq:lyapunov_perfect_sensing_large_k} can be replaced by the \emph{effective} strategy $(1-\epsilon)\bm{\pi}_{e}+\epsilon\bm{\pi}_{e'}$. We call this generalized Lyapunov exponent $\Lambda^{(a,\epsilon)}_{k\rightarrow{}\infty}$ when $k\rightarrow\infty$ and as with the case of perfect sensing, we can plug the effective strategy into Eq. \ref{eq:lyapunov_approx} when $k$ is finite in order to find $\Lambda^{(a,\epsilon)}_{k}$. However, introducing a perceived environmental state which will fluctuate even in a constant environment brings up the same difficulties in analysis which made the adiabatic assumption necessary. In order for us to include a non-zero environmental sensing error probability, we make the additional assumption that the environmental state perceived by an organism is decided at the onset of an environmental change and is maintained until the environment again changes. 

We can also use Eq. \ref{eq:lyapunov_perfect_sensing_large_k} to write down the Lyapunov exponent for a population of organisms which do not adjust their phenotype switching rates according to the environmental state. Because the phenotype switching rates are independent of the environment, there is no relaxation cost and we can write the Lyapunov exponent simply as
\begin{equation}\label{eq:lyapunov_no_sensing}
\Lambda^{(\eta)}=\sum_{e\in\mathcal{E}}p(e)\log(\mathbf{w}_{e}\bm{\pi})
\end{equation}
where $\eta$ denotes an informationless channel. Note that the adiabatic limit is no longer required for analysis, as $\bm{\pi}$ is independent of the environment. The informationless Lyapunov exponent $\Lambda^{(\eta)}$ corresponds to $\Lambda^{(a,\epsilon=1/2)}$, where the mutual information between the perceived environmental and actual environmental states is zero.

\section{Allowing for environment-dependent phenotype relaxation rates}\label{section:appendixb}

As discussed in the main text, when $\epsilon\in(0,1/2)$ the strategy must be replaced by an effective strategy. We are able to do this because the perceived environmental state of each individual is independent from that of each other individual. However, our use of an effective strategy as in Eqs. \ref{eq:eff_strat1}-\ref{eq:eff_strat4} would be inappropriate if there was instead a global environmental cue with a certain probability of accurately representing the true environmental state (see \cite{donaldson2013unreliable} for further discussion of these separate error sources).

The effective strategy can be derived as follows. For constant environmental state $e_{1}$, the probability distribution over phenotypes at time step $m$ is related to that at time step $m+1$ according to

\begin{equation}
\textbf{p}_{\bm{\phi},m+1}=\Big((1-\epsilon)\mathbf{P}_{\tilde{e}_{1}}+\epsilon\mathbf{P}_{\tilde{e}_{2}}\Big)\mathbf{p}_{\bm{\phi},m}
\end{equation}

while in constant environmental state $e_{2}$, this relationship becomes 

\begin{equation}
\textbf{p}_{\bm{\phi},m+1}=\Big(\epsilon\mathbf{P}_{\tilde{e}_{1}}+(1-\epsilon)\mathbf{P}_{\tilde{e}_{2}}\Big)\mathbf{p}_{\bm{\phi},m}.
\end{equation}

We have not only assumed that environmental observations are independent between individuals, but also that successive observations by the same individual are also independent. Then, the overall effective strategy can be written according to the first eigenvectors of $(1-\epsilon)\mathbf{P}_{\tilde{e}_{1}}+\epsilon\mathbf{P}_{\tilde{e}_{2}}$ and $\epsilon\mathbf{P}_{\tilde{e}_{1}}+(1-\epsilon)\mathbf{P}_{\tilde{e}_{2}}$, yielding 

\begin{align}\label{eq:general_eff_strat1}
\rho_{\phi_{1},e_{1}}(\epsilon)&=\frac{(1-\epsilon)\omega_{\tilde{e}_{1}}+\epsilon\omega_{\tilde{e}_{2}}}{(1-\epsilon)\chi_{\tilde{e}_{1}}+(1-\epsilon)\omega_{\tilde{e}_{1}}+\epsilon\chi_{\tilde{e}_{2}}+\epsilon\omega_{\tilde{e}_{2}}}\\\label{eq:general_eff_strat2}
\rho_{\phi_{2},e_{1}}(\epsilon)&=1-\rho_{\phi_{1},e_{1}}(\epsilon)\\\label{eq:general_eff_strat3}
\rho_{\phi_{1},e_{2}}(\epsilon)&=\frac{\epsilon\omega_{\tilde{e}_{1}}+(1-\epsilon)\omega_{\tilde{e}_{2}}}{\epsilon\chi_{\tilde{e}_{1}}+\epsilon\omega_{\tilde{e}_{1}}+(1-\epsilon)\chi_{\tilde{e}_{2}}+(1-\epsilon)\omega_{\tilde{e}_{2}}}\\\label{eq:general_eff_strat4}
\rho_{\phi_{2},e_{2}}(\epsilon)&=1-\rho_{\phi_{1},e_{2}}(\epsilon).
\end{align}

In this general case, the phenotype relaxation rates in each environment are not necessarily the same and are determined by the second eigenvalues of these two matrices, so that

\begin{align}\nonumber
k_{e_{1}}&=-\log\Big(1-(1-\epsilon)\chi_{\tilde{e}_{1}}-(1-\epsilon)\omega_{\tilde{e}_{1}}-\epsilon\chi_{\tilde{e}_{2}}-\epsilon\omega_{\tilde{e}_{2}}\Big)\\\nonumber
&=-\log\Big((1-\epsilon)(1-\chi_{\tilde{e}_{1}}-\omega_{\tilde{e}_{1}})+\epsilon(1-\chi_{\tilde{e}_{2}}-\omega_{\tilde{e}_{2}})\Big)\\\label{eq:k_e1}
&=-\log\Big((1-\epsilon)\exp(-k_{\tilde{e}_{1}})+\epsilon{}\exp(-k_{\tilde{e}_{2}})\Big)\\\nonumber
k_{e_{2}}&=-\log\Big(1-\epsilon\chi_{\tilde{e}_{1}}-\epsilon\omega_{\tilde{e}_{1}}-(1-\epsilon)\chi_{\tilde{e}_{2}}-(1-\epsilon)\omega_{\tilde{e}_{2}}\Big)\\\nonumber
&=-\log\Big(\epsilon(1-\chi_{\tilde{e}_{1}}-\omega_{\tilde{e}_{1}})+(1-\epsilon)(1-\chi_{\tilde{e}_{2}}-\omega_{\tilde{e}_{2}})\Big)\\\label{eq:k_e2}
&=-\log\Big(\epsilon{}\exp(-k_{\tilde{e}_{1}})+(1-\epsilon)\exp(-k_{\tilde{e}_{2}})\Big).
\end{align}

If $k_{\tilde{e}_{1}}=k_{\tilde{e}_{2}}$, it is clear that $k_{e_{1}}=k_{e_{2}}$ also holds. If $k_{\tilde{e}_{1}}\neq{}k_{\tilde{e}_{2}}$, then Eqs. \ref{eq:k_e1}-\ref{eq:k_e2} should be used in place of $k$ in the derivations contained in Appendix \ref{section:appendixa}.

When $k_{\tilde{e}_{1}}=k_{\tilde{e}_{2}}$, so that $\chi_{\tilde{e}_{1}}+\omega_{\tilde{e}_{1}}=\chi_{\tilde{e}_{2}}+\omega_{\tilde{e}_{2}}$, Eqs. \ref{eq:general_eff_strat1}-\ref{eq:general_eff_strat4} reduce to Eqs. \ref{eq:eff_strat1}-\ref{eq:eff_strat4}. In the limit of large $k_{\tilde{e}_{1}}$ and $k_{\tilde{e}_{2}}$ where the two rates are not necessarily exactly the same, an effective strategy as in Eqs. \ref{eq:eff_strat1}-\ref{eq:eff_strat4} is approximately accurate. In this case, the results do not change for unequal $k_{\tilde{e}_{1}}$ and $k_{\tilde{e}_{2}}$, as both are ``approximately" infinite. However, when $k_{\tilde{e}_{1}}$ and $k_{\tilde{e}_{2}}$ are relatively small, the effective strategies shown in Eqs. \ref{eq:general_eff_strat1}-\ref{eq:general_eff_strat4} must be substituted in. 

\begin{figure}[ht!]
    \centering
    \includegraphics[width=.75\textwidth]{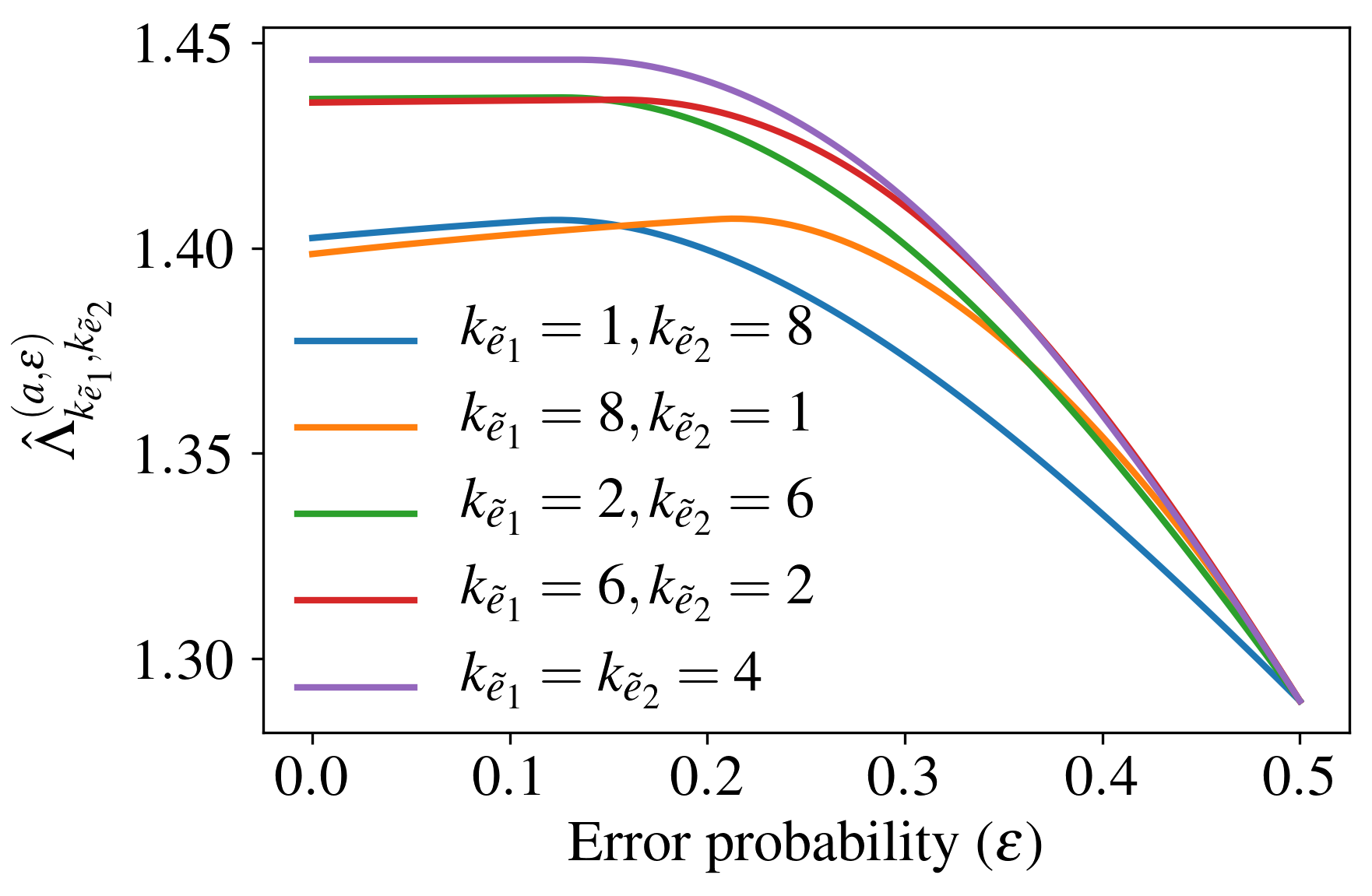}
    \caption{{\bf Optimal Lyapunov exponents when $\mathbf{k_{\tilde{e}_{1}}\neq{}k_{\tilde{e}_{2}}}$.} We numerically optimized Lyapunov exponents for a number of cases where $k_{\tilde{e}_{1}}$ and $k_{\tilde{e}_{2}}$ are not necessarily equal. The purple curve with $k_{\tilde{e}_{1}}=k_{\tilde{e}_{2}}=4$ is the special case examined in the main text. In some other cases, a critical error probability $\epsilon_{c_{1}}$ exists, such as when $k_{\tilde{e}_{1}}=2$ and $k_{\tilde{e}_{2}}=6$ or when $k_{\tilde{e}_{1}}=6$ and $k_{\tilde{e}_{2}}=2$. Interestingly, when $k_{\tilde{e}_{1}}=1$ and $k_{\tilde{e}_{2}}=8$ or $k_{\tilde{e}_{1}}=8$ and $k_{\tilde{e}_{2}}=1$, the Lyapunov exponent increases with $\epsilon$ and reaches a maximum before decreasing. Model parameters are fixed at $w_{\phi_{1}e_{1}}=5.0$, $w_{\phi_{2}e_{1}}=0.01$, $w_{\phi_{1}e_{2}}=1.0$, $w_{\phi_{2}e_{2}}=10.0$, $\alpha=0.2$, and $\beta=0.25$. This figure was created in a Python 3.7 Jupyter Notebook \protect\cite{kluyver2016jupyter} with Matplotlib 3.1.1 \protect\cite{hunter2007matplotlib}.}
    \label{fig:appendix_different_k}
\end{figure}

Using numerical optimization, we find that even with distinct phenotype relaxation rates, there can still be a critical error probability (Fig. \ref{fig:appendix_different_k}). In several cases, we found similar behavior to the case of identical phenotype switching rates, where no change in the optimal Lyapunov exponent occurs below a critical error probability. However, in other cases the optimal Lyapunov exponent increased to a maximal value before decreasing. This indicates that even without considering costs, a non-zero optimal sensing error probability can exist.  

\section{Optimal environment-agnostic phenotype switching strategies}\label{section:appendixc}

Here we describe optimal strategies in the simplest case where individuals cannot sense the environmental state. We set $\epsilon$ to $1/2$, so that $k$ becomes irrelevant, as described earlier. Then Eq. \ref{eq:full_lyapunov} becomes $\Lambda^{(\eta)}$ (Eq. \ref{eq:lyapunov_no_sensing}). As this population  cannot sense the environment in any useful way, the phenotype distribution is independent of the environment and is written as $\pi_{\phi_{1}}$ and $\pi_{\phi_{2}}$ where $\pi_{\phi_{1}}+\pi_{\phi_{2}}=1$. Setting the derivative of $\Lambda^{(\eta)}$ with respect to $\pi_{\phi_{1}}$ equal to zero, we find the optimal phenotypic distribution to be
\begin{align}\label{eq:no_sensing_solution1}
\hat{\pi}_{\phi_{1}}&=\min\Bigg[\max\Bigg[p(e_{1})\bigg(1-\frac{w_{\phi_{1}e_{2}}}{w_{\phi_{2}e_{2}}}\bigg)^{-1}+p(e_{2})\bigg(1-\frac{w_{\phi_{1}e_{1}}}{w_{\phi_{2}e_{1}}}\bigg)^{-1},0\Bigg],1\Bigg]\\\label{eq:no_sensing_solution2}
\hat{\pi}_{\phi_{2}}&=1-\hat{\pi}_{\phi_{1}}.
\end{align}
This is exactly the solution found by Rivoire and Leibler (Appendix D, Eq. 67) \cite{rivoire2011value} to a model previously explored in Donaldson-Matasci, Bergstrom, and Lachmann \cite{donaldson2010fitness}. Unsurprisingly, the optimal phenotype distribution depends on both the environmental switching rates, by means of the environmental state probabilities, and the fitness of each phenotype in each environment. If we approach the case where the ``incorrect'' phenotype for each environment causes all individuals of that phenotype to die, that is $w_{\phi_{2},e_{1}},w_{\phi_{1},e_{2}}\rightarrow{}0$, then the optimal strategy approaches proportional betting where $\hat{\pi}_{\phi_{1}}\rightarrow{}p(e_{1})$ and $\hat{\pi}_{\phi_{2}}\rightarrow{}p(e_{2})$.

What then is the optimal long-term growth rate? Plugging the optimal distribution from Eqs. \ref{eq:no_sensing_solution1} \& \ref{eq:no_sensing_solution2} into Eq. \ref{eq:lyapunov_no_sensing}, we have
\begin{equation}
\hat{\Lambda}^{\eta}=\sum_{e\in\mathcal{E}}p(e)\log\bigg(\frac{\det\mathbf{W}}{\Delta{}w_{e'}}\bigg)-H(E)
\end{equation}
where $H(E)$ is the entropy of the environment. This result was previously reported by Donaldson-Matasci, Bergstrom, and Lachmann \cite{donaldson2010fitness}. This holds true when bet-hedging is optimal, whereas when pure strategies are optimal the optimal growth rate can be written as
\begin{equation}
\hat{\Lambda}^{\eta}=\sum_{e\in\mathcal{E}}p(e)\log\bigg(\frac{w_{\hat{\phi},e}}{p(e)}\bigg)-H(E)
\end{equation}
where $w_{\hat{\phi},e}$ is the fitness of the sole expressed phenotype $\hat{\phi}$ in environment $e$. For fixed fitness values, the optimal Lyapunov exponent is minimized when $\alpha/\beta=\Delta{}w_{e_{1}}/\Delta{}w_{e_{2}}$, so that when $\Delta{}w_{e_{1}}=\Delta{}w_{e_{2}}$ the minimal optimal Lyapunov exponent is achieved when environmental entropy is maximized.

\bibliography{bibliography}

\end{document}